\documentclass[twocolumn,showpacs,aps,prd,letterpaper]{revtex4-1}
\usepackage{graphicx}
\usepackage{dcolumn}
\usepackage{amsmath}
\usepackage{epsfig}
\RequirePackage{xspace}

\usepackage{relsize}
\begin{document}

\title{Sudakov Factor in the Deep Inelastic Scattering \\ of a Current off a Large Nucleus}
\author{Ya-ping Xie}\email{xieyaping@impcas.ac.cn}\author{Xurong Chen}\email{xchen@impcas.ac.cn}
\affiliation{Institute of Modern Physics, Chinese Academy of
Sciences, Lanzhou 730000, China}
\begin{abstract}
We consider a gedanken experiment of the scattering of a current
$j=-\frac{1}{4}F^i_{\mu\nu}F^i_{\mu\nu}$  off a large nucleus to
study the gluon saturation at the small-$x$ limit and compute the
Sudakov factor of this process through a one-loop calculation. The
differential cross section is expressed in term of the Sudakov
resummation, in which the collinear and the rapidity divergences
are subtracted. We also discuss how to probe the
Weizs\"{a}cker-Williams (WW) gluon distribution in the process of
photon pair production in the $pA$ collisions.
\end{abstract}
 %14.80.Bn
\maketitle
\section{Introduction}
Saturation physics\cite{Gribov:1984tu, Mueller:1989st} has been
one of the most interesting topics in high energy nuclear physics
corresponding the small-$x$ physics. It describes the rapid rise
of parton distributions at very high energy
\cite{Balitsky:1978ic,Kuraev:1977fs}. In addition, by including
the non-linear evolution\cite{Mueller:1985wy, Balitsky:1995ub,
JalilianMarian:1997jx} dynamics when the gluon distribution
becomes order $1/\alpha_s$, parton distributions start to saturate
as the scattering amplitude approaches unitarity. There are many
phenomenological models to describe the dynamics inside the
nucleuses in small-$x$ physics, the semi-classic model named color
glass condensate (CGC) has been widely used to describe the
dynamics inside the nucleus in the small-$x$
physics\cite{Iancu:2000hn}.   \\
\indent  The Sudakov resummation  has been widely used in various
high energy physics processes\cite{Sudakov:1954sw, Collins:1984kg,
Davies:1984sp}, in the meantime, the small-$x$ logarithms
 $\alpha_s\ln{1/x_g}$ are also important, and normally resummed through small-$x$
 evolution equations. It has
been demonstrated that Sudakov type large logarithms and small-$x$
type logarithms can be resummed independently in various physics
processes, for example, the Higgs production and dijet in $pA$
collisions\cite{Mueller:2012uf,
Mueller:2013uf}.\\
 \indent The so-called Weiz\"{a}cker-Williams (WW) gluon
distribution\cite{Collins:1981uw, McLerran:1993ni, Ji:2005nu,
Dominguez:2011gc}, which is the genuine gluon distribution, only
appears in the observables if the initial or the final state
interaction is absent in the inelastic scattering of the gluonic
current on a nucleus target. For example, in the photon pair
production in $pA$ collisions, where the final interaction is
absent, the WW gluon
distribution appears.\\
\indent The objective of this calculation is to investigate the
DIS of the gluonic current off a large nucleus and the photon pair
production in $pA$ collisions. The common feature of these two
processes is that they are both involving the so-called WW gluon
distribution at small-$x$ limit. This paper can be considered as a
supplementary study of the Ref.~\cite{Mueller:2013uf}.\\
\indent The paper is organized into five sections, we start with
introduction, The section II is about the leading order of DIS of
gluonic current off a large nucleus. The details of calculation of
the Sudakov factor will be presented in section III, we discuss
how to probe the WW gluon distribution in the process of photon
pair production in $pA$ collisions in section IV, in section V, we
conclude with discussions on Sudakov resummmation and WW gluon
distribution.

\section{ Leading order in DIS of Current off a large nucleus}
\begin{figure}[ht]
\begin{center}
\includegraphics[width=5cm]{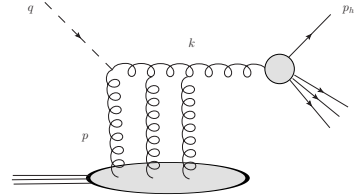}
\caption{ Feynman diagram at leading order of deep inelastic
scattering of the current off a large nucleus. }\label{fig1}
\end{center}
\end{figure}

 First, we consider a gedanken experiment of deep inelastic scattering
(DIS) of the gluonic current of
$j(x)=-\frac{1}{4}F^{i}_{\mu\nu}(x)F^{i}_{\mu\nu}(x)$ off a large
nucleus at the leading order\cite{Mueller:1989st,
Kovchegov:1998bi}, where $F^i_{\mu\nu}(x)$ is the QCD field
strength tensor. This current is chosen because it is easy to
study the WW gluon distribution in this process. The feynman
diagram of DIS of the gluonic current off a large nucleus is in
Fig.~\ref{fig1}, the momentum of the gluonic current is $q$, we
can treat the current as a
  scalar particle, the  momentum square of the
  current satisfies $q^2=-Q^2$, and the virtual mass of the current is
  $M=iQ$, the  notation of $q$ in the light cone gauge is
 \begin{equation}
q^\mu=\left(q^+,\frac{-Q^2}{2q^+},0\right).
\end{equation}
 We define $k$ as momentum of the final state produced gluon(the
 horizontal gluon in Fig.~\ref{fig1}),
  $P$ as momentum of the large nucleus, $p$ as momentum of the outgoing
gluons(vertical gluons in Fig.~\ref{fig1})
 from the target nucleus. In this calculation, we assume that the plus component of $q$
 and minus component of $P$ both are large.
We define  $p_h$  as momentum of the final state
 hadron, $y$ as the rapidity of final state hadron, and $z_1=p_h^+/k^+$.
 We can calculate the differential cross section by
transverse momentum dependent(TMD) factorization\cite{Ji:2005nu,
Dominguez:2011wm}, it reads
\begin{eqnarray}
\frac{\text{d}\sigma^{jA\rightarrow h+X}_{\textrm{LO}}}{
\sigma_0\text{d} y \text{d}^2 p_{h\perp}}
&=&\frac{D_{h/g}(z_1)}{z^2_1} \int\frac{d^2x_\perp
d^2x^{\prime}_\perp}{(2\pi)^2}\,e^{-ik_\perp\cdot(x_\perp-x^{\prime}_\perp)}\notag \\
&&\times S^{WW}(x_\perp,x^\prime_\perp),
\end{eqnarray}
where $S^{ww}(x_\perp, x^\prime_\perp)$ is the WW gluon
distribution in the coordinate space, it is defined as
\begin{eqnarray}
S&&^{WW}(x_\perp,x^\prime_\perp)\notag \\&&=-
\langle\text{Tr}[\partial^iU(x_\perp)]U^\dagger(x^{\prime}_%
\perp)[\partial^iU(x^{\prime}_\perp)]U^\dagger(x_\perp)
\rangle_{x_g}, \label{sww}
\end{eqnarray}
 the fundamental Wilson line is defined as
\begin{equation}
U(x_\perp)=\mathcal{P}\exp\left\{ig_S\int_{-\infty}^{+\infty} \text{d}%
x^+\,T^cA_c^-(x^+,x_\perp)\right\} \ ,
\end{equation}
where $A_c^-(x^+,x_\perp)$ is gluon field solution of Yang-Mill
equation. There is a $\delta(z_1-p_h^+/k^+)$, which has been
integrated by $z_1$. The $\sigma_0$ is the leading-order of the
gluon production, and it reads $\sigma_0 \propto
1/4g^2\left(1-\epsilon\right)$, where $\epsilon=(4-D)/2$.
  The kinematics of DIS of the gluonic current off a large
  nucleus target at leading-order satisfies
\begin{equation}
  x_gs=Q^2,\label{xg}
\end{equation}
where $s=2P\cdot q$, and $x_g$ is longitudinal momentum fraction
of the outgoing gluons to the nucleus target, we assume that
$Q^2\gg k_\perp^2$, but keep $x_g\ll 1$. In addition, throughout
this paper, we use leading power approximation, therefore, we
neglect higher order power correction of order $k^2_\perp/Q^2$.
\section{Sudakov factor in DIS of current off a large nucleus }
Now, we consider one-loop order of the process of DIS of current
off a large nucleus, with the help of formalism of dipole
model\cite{Mueller:1999wm}, it is more convenient to do the
calculation in coordinate space. In order to calculate the
amplitude in coordinate space, some types of splitting functions
are introduced at first, such as $\Psi_{g\to gg}\left(\xi,
u_\perp\right)$ , $\Psi_{g\to q\bar{q}}\left(\xi, u_\perp\right)$
and $\Psi_{j\to gg}\left(\xi,
u_\perp\right)$\cite{Mueller:2013uf, Dominguez:2011wm}.\\
 \indent The $j \rightarrow gg$ splitting function in momentum space
and coordinate space are
\begin{eqnarray}
\Psi _{j \rightarrow gg}\left( \xi ,k_{\perp }\right) &=&\sqrt{\frac{1}{%
2\xi (1-\xi )k^{+}}}\frac{1}{k_{\perp }^{2}+\xi (1-\xi
)Q^{2}}\notag \\
&&\times\left( \frac{1}{2}k_{\perp }^{2}\epsilon _{\perp
}^{(1)}\cdot \epsilon _{\perp }^{(2)}-k_{\perp }\cdot \epsilon
_{\perp }^{(1)}k_{\perp }\cdot \epsilon _{\perp }^{(2)}\right),
\notag \\
\label{sfm}
\end{eqnarray}
and
\begin{eqnarray}
\Psi _{j \rightarrow gg}\left( \xi ,u_{\perp }\right)
&=&-\sqrt{\frac{1}{ 2\xi (1-\xi )k^{+}}}\frac{2\pi }{u_{\perp
}^{2}}K(\epsilon _{f}u_{\perp })\notag \\ &&\times\left(
\frac{1}{2}\epsilon _{\perp }^{(1)}\cdot \epsilon _{\perp
}^{(2)}-\frac{1}{u_{\perp }^{2}}u_{\perp }\cdot \epsilon _{\perp
}^{(1)}u_{\perp }\cdot \epsilon _{\perp }^{(2)}\right) ,\notag \\
\end{eqnarray}
respectively, here $\xi$  and $1-\xi$ are fractions of the
longitudinal momentum of the radiating gluons,
$\epsilon_\perp^{(1)}$ and $\epsilon_\perp^{(2)}$ are
polarizations of the radiating gluons, and $K(\epsilon
_{f}u_{\perp })$ is defined as
\begin{eqnarray}
K(\epsilon _{f}u_{\perp }) = 2\epsilon _{f}u_{\perp }\text{K}
_{1}(\epsilon _{f}u_{\perp })+\epsilon _{f}^{2}u_{\perp }^{2}
\text{K}_{0}(\epsilon _{f}u_{\perp }),
\end{eqnarray}
where  $\epsilon _{f}^{2}=\xi (1-\xi )Q^{2}$ and $K_{0,1} $
 are modified Bessel functions.\\
 \indent   Because the scattering energy in the collisions is very high, the interaction time
of multiply scattering is so short that the radiating gluon is
either before or after the multiply scattering.
\begin{figure}[ht]
\begin{center}
\includegraphics[width=5cm]{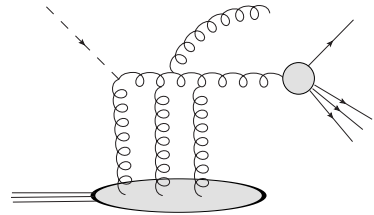}\hspace{1.0cm}
\caption{Radiating gluon happens between the multiply scattering,
this process is neglected because the time for multiply scattering
between the produced gluon and nucleus target is too short in the
high energy limit.}\label{fig1b}
\end{center}
\end{figure}
 The process that
radiating gluons happens between the multiply scattering as
illustrated in Fig.~\ref{fig1b} can be neglected. On the other
hand, this diagram will be important when we have very large
target, while the scattering energy is not high.\\
 \indent  The
feynman diagrams of DIS of gluonic current
  off a large nucleus at one-loop order are
described in Fig.~\ref{fig2} and Fig.~\ref{fig3}. Fig.~\ref{fig2}
depicts the real feynman diagrams, Fig.~\ref{fig3} illstrates the
virtual feynman diagrams. We firstly study the real diagrams, we
can see that the radiating gluon is after the multiply scattering
in graph $(a)$ in Fig.~\ref{fig2}, the radiating gluon is before
the multiply scattering in graph $(b)$ in Fig.~\ref{fig2}. The
differences between them result different contributions in leading
power approximation. As discussion in Ref.~\cite{Mueller:2013uf},
in the case of $k^2_\perp \ll Q^2$, from Eq.~(\ref{sfm}), we can
see that the value of splitting function $\Psi_{j\to gg}(\xi,
k_\perp)$ is suppressed when $\xi\neq 1$. We can see that the
graph $(b)$ in Fig.~\ref{fig2} which
 radiating gluon is before the multiply scattering is leading power suppressed,
 because the splitting function is $\Psi_{j\to gg}(\xi, k_\perp)$.
 Thus, the contributions of square of graph
$(b)$ and the interference of graph $(a)$ and $(b)$ in
Fig.~\ref{fig2} are power suppressed, the contribution of the
square of graph $(a)$ in Fig.~\ref{fig2}
 is the only real leading power contribution. \\
 \begin{widetext}
\indent The differential cross section of square of graph $(a)$ in
Fig.~\ref{fig2} can be cast into
\begin{eqnarray}
\frac{\text{d}\sigma_a^{jA\to h X}}{\sigma_0\text{d} y\text{d}^2 p_{h\perp}}&=&k^+%
\alpha_s N_c
\int\text{d}z^\prime_1\int_\tau^1\frac{d\xi}{z^{\prime
2}_1}D(z^\prime_1)
\delta(z^\prime_1\xi-p^+_h/k^+)\int \frac{\text{d}%
^{2}x_{\perp}}{(2\pi)^{2}}\frac{\text{d}^{2}x_{\perp}^{\prime }}{(2\pi )^{2}}%
\frac{\text{d}^{2}b_{\perp}}{(2\pi)^{2}}\frac{\text{d}^{2}b^\prime_{\perp}}{(2\pi)^{2}}
\int\text{d}^{2} l_{2\perp}\notag \\
&&\times e^{-il_{1\perp }\cdot(x_{\perp}-x^{\prime
}_{\perp})}e^{-il_{2\perp }\cdot(b_{\perp}-b^{\prime}_{\perp})}
\sum \Psi^{\ast}_{g\to g g}\left(\xi,
u^{\prime}_\perp\right)\Psi_{g\to gg}\left(\xi, u_\perp\right)
S^{WW}(v_\perp,v^\prime_\perp),
\end{eqnarray}
 where $l_{1}$ and $l_{2}$ are momentum of the outgoing gluons
of graph $(a)$ in Fig.~\ref{fig2}. The fraction $\xi$ is defined
as $\xi=l_{1}^+/k^+$, and $z^\prime_1$ is defined as
$z^\prime_1=p_{h}^+/l_1^+$. The range of variable
 $\tau$ is $z_1<\tau <1$, where $z_1=p_h^+/k^+$.
\begin{figure}[h]
\begin{center}
\includegraphics[width=4in]{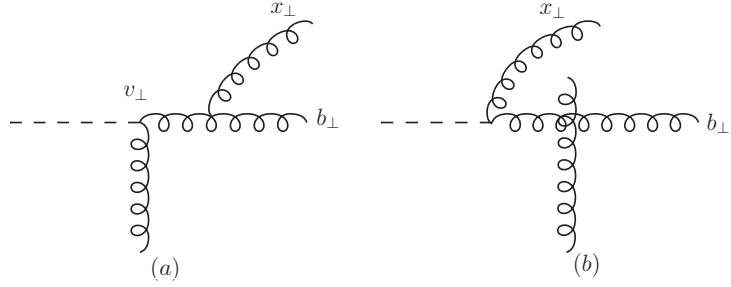}
\caption{The real diagrams of deep inelastic scattering of the
current off a large nucleus at one-loop order.}\label{fig2}
\end{center}
\end{figure}
 The coordinate variables are defined as $u_\perp=x_\perp-b_\perp$,
$v_\perp=\xi x_\perp+(1-\xi)b_\perp$ and
$u^\prime_\perp=x^\prime_\perp-b^\prime_\perp$,
$v^\prime_\perp=\xi x^\prime_\perp+(1-\xi)b^\prime_\perp$, using
these relationships, we can change the integral variables in the
phase space integral, after some algebraic derivations, we get
\begin{eqnarray}
\frac{\text{d}\sigma_a^{jA\to h X}}{\sigma_0\text{d} y\text{d}^2 p_{h\perp}}&=&k^+%
\alpha_s N_c
\int_{\tau}^1\frac{d\xi}{z^2_1}D(z_1/\xi)\frac{1}{\xi}
 \int \frac{\text{d}
^{2}u_{\perp}}{(2\pi)^{2}}
\frac{\text{d}^{2}v_{\perp}}{(2\pi)^{2}}\frac{\text{d}^{2}v^\prime_{\perp}}
{(2\pi)^{2}} \notag \\
&&\times e^{-ik_{\perp }\cdot(v_{\perp}-v^{\prime }_{\perp})} \sum
\Psi^{\ast}_{g\to g g}\left(\xi, u^{\prime}_\perp\right)\Psi_{g\to
gg}\left(\xi, u_\perp\right)
S^{WW}(v_\perp,v^\prime_\perp),\label{beforsplitting}
\end{eqnarray}
 where
$u_\perp^\prime=u_\perp-\frac{1}{\xi}(v_\perp-v^\prime_\perp)=u_\perp-(
x_\perp-x^\prime_\perp)$, and we use $l_{1\perp}/k_\perp\approx
l_{1}^+/k^+=\xi$. Substituting the sum of $\Psi_{g\to gg}(\xi,
u_\perp)$\cite{Dominguez:2011wm}
\begin{equation}
\sum \Psi_{g\to gg}^*(\xi,u_\perp^{\prime})\Psi_{g\to
gg}(\xi,u_\perp)=(2\pi)^2
\frac{4}{k^+}\left[\frac{\xi}{1-\xi}+\frac{1-\xi}{\xi}+\xi(1-\xi)\right]\frac{u_\perp^{\prime}\cdot
u_\perp}{u_\perp^{\prime 2} u_\perp^{ 2}}
\end{equation}
into Eq.~(\ref{beforsplitting}), we get
\begin{eqnarray}
\frac{\text{d}\sigma_a^{jA\to h X}}{\sigma_0\text{d} y\text{d}^2
p_{h\perp}}&=&4\alpha_s
N_c\int\frac{\text{d}^{2}v_{\perp}\text{d}^{2}v^\prime_{\perp}}{(2\pi)^{2}}
e^{-ik_{\perp }\cdot(v_{\perp}-v^{\prime}_{\perp})}S^{WW}(v_\perp,v^\prime_\perp)\notag \\
&&\times\int_{\tau}^1\frac{d\xi}{z^2_1}D(z_1/\xi) \int
\frac{\text{d}^2l_{2\perp}}{(2\pi)^2}e^{-il_{2\perp}\cdot
\frac{1}{\xi}(v_\perp-v^\prime_\perp)}\frac{1}{l_{2\perp}^2}\frac{1}{\xi}
 \left[\frac{1}{1-\xi}+\frac{1-\xi}{\xi}+\xi(1-\xi)\right].
\label{twoline}
\end{eqnarray}
\end{widetext}
We can see that the first line of Eq.~(\ref{twoline}) is
proportional to the leading power differential cross section,
which should be factorized out. The integral of the second line of
Eq.~(\ref{twoline}) contains various types of divergences. In
order to obtain the Sudakov logarithms terms, we should subtract
these divergences. These divergences include rapidity divergence,
collinear divergence, and other divergences. The rapidity
divergence associates with the WW gluon distribution of the large
nucleus target\cite{Dominguez:2011gc}, and the collinear
divergence associates with the fragmentation function of the final
hadron\cite{Chirilli:2011km}, other divergences should be
cancelled by virtual loops. It is necessary to subtract these
divergences in order to obtain the Sudakov logarithms terms.
Firstly, we can subtract the collinear divergence using the plus
function.
 We can rewrite the second
line of Eq.~(\ref{twoline}) as
\begin{eqnarray}
 &&\int_{\tau}^1\frac{d\xi}{ z^2_1}D(z_1/\xi) \frac{\text{d}^2l_{2\perp}}{(2\pi)^2}e^{-il_{2\perp}\cdot
R_\perp/\xi}\notag \notag
\\&&\times\frac{1}{l_{2\perp}^2\xi}\left[\frac{1}{(1-\xi)_+}+\frac{1-\xi}{\xi}+\xi(1-\xi)\right]
\notag
\\&&+\frac{D(z_1)}{z^2_1}\int\frac{\text{d}^2l_{1\perp}}{(2\pi)^2}e^{-il_{2\perp}\cdot
R_\perp}\frac{1}{l_{2\perp}^2}\int^1_0\text{d}\xi\frac{1}{1-\xi},\label{subtract}
\end{eqnarray}
 where $R_\perp=(v_\perp-v^\prime_\perp)$, the first
term of Eq.~(\ref{subtract}) can be interpreted as part of
renormalization of the fragmentation function of final state
hadron.
 We take $l_{2\perp}^\prime=l_{2\perp}/\xi$, and integrate it by $l_{2\perp}^\prime$, the integral
of the first line is proportional to
\begin{eqnarray}
    \frac{1}{4\pi}\frac{1}{\xi}\left[\frac{\xi}{(1-\xi)}_++\frac{1-\xi}{\xi}+\xi(1-\xi)\right]
    \left(-\frac{1}{\epsilon}+\ln\frac{c_0^2}{\mu^2R^2_\perp}\right),\notag \\
    \label{partofRF}
\end{eqnarray}
where we use $\overline{MS}$ scheme, it is part of the splitting
function. Then, we are going to calculate the integral of the
second line of Eq.~(\ref{subtract}). According to momentum energy
conversation, we can get the kinematics of graph $(a)$ in
Fig.~\ref{fig2} as follows
\begin{equation}
  x^\prime_gs=\frac{l^2_{2\perp}}{(1-\xi)}+\frac{l^{2}_{1\perp}}{\xi}.
\end{equation}
In $\xi\rightarrow 1$ limit, we get
\begin{equation}
  \xi<1-\frac{l^2_{2\perp}}{s}.
\end{equation}
 Now, the last integral of the
second line of Eq.~(\ref{subtract}) can be written as
\begin{eqnarray}
  &&\int^{1-\frac{l^2_{2\perp}}{s}}_0\frac{1}{1-\xi}=
\ln\frac{s}{l_{2\perp}^2}=
\ln\frac{s}{Q^2}+\ln\frac{Q^2}{l_{2\perp}^2}\notag \\
&=&\ln\frac{1}{x_g}+\ln\frac{Q^2}{l^2_{2\perp}}.\label{twolog}
\end{eqnarray}
Back to Eq.~(\ref{xg}), we get $x_g \to 0$, as $s\to \infty$,
$\ln1/x_g$ is divergent. In small-$x$ physics, the product of
$\alpha_s$ and $\ln1/x_g$ is resummed through the small-$x$
evolution equation of the WW gluon distribution. Thus the first
logarithm term of Eq.~(\ref{twolog}) should be separated out from
Sudakov resummation, and it should be absorbed into the
renormalization of the WW gluon
distribution\cite{Dominguez:2011gc}. The evolution equation of the
WW gluon distribution is
\begin{eqnarray}
\frac{\partial}{\partial\ln 1/x_g}S^{WW}(x_\perp,x^\prime_\perp)
 =\int {\bf K}_{\rm DMMX}\otimes
 S^{WW}(x_\perp,x^\prime_\perp),\notag \\
 \label{wwe}
\end{eqnarray}
where  ${\bf K}_{\rm DMMX}$ is the kernel of the small-$x$
evolution equation.
 Besides these two divergences, the integral of
the second logarithm term of Eq.~(\ref{twolog}) contains other
divergences, which can be cancelled by contributions of virtual
loops, we can write the integral of the second logarithm term of
Eq.~(\ref{twolog}) as
\begin{equation}
\mu^{2\epsilon}\int\frac{\text{d}^{2-2\epsilon}
l_{2\perp}}{(2\pi)^{2-2\epsilon}}e^{-il_{2\perp}\cdot
R_\perp}\frac{1}{l_{2\perp}^{ 2}}\ln\frac{Q^2}{l^2_{2\perp}},
\label{realsudakov}
\end{equation}
where the integral dimension has been changed from $2$ to
$2-2\epsilon$. Using the formulas in Appendix of
Ref.~\cite{Mueller:2013uf}, we obtain the contribution of square
of graph $(a)$ in Fig.~\ref{fig2}
\begin{eqnarray}
\frac{\alpha_sN_c}{\pi}\left(\frac{1}{\epsilon^2}-\frac{1}{\epsilon}\ln\frac{Q^2}{\mu^2}
+\frac{1}{2}\ln^2\frac{Q^2}{\mu^2}-\frac{1}{2}\ln^2\frac{Q^2
R_\perp^2}{c_0^2}-\frac{\pi^2}{12}\right), \notag \\
\label{rc}
\end{eqnarray}
where $ c_0=2e^{-\gamma_E}$,  $\gamma_E\approx0.5772$ is Euler
constant.\\
\indent There are three kinds of virtual graphs in the DIS of
current off a large nucleus as described in Fig.~\ref{fig3}, the
graph $(c)$ in Fig.~\ref{fig3} is the gluon self-energy diagram,
the graph $(e)$ in Fig.~\ref{fig3} is the quark self-energy
diagram. We begin with the calculation of the virtual graph  $(d)$
in Fig.~\ref{fig3}, and find
\begin{widetext}
\begin{eqnarray}
 \frac{\text{d}\sigma_{d}^{jA\to h X}}{\sigma_0\text{d}
y\text{d}^2 p_{h\perp}}&&=-ik^+\alpha_s \frac{D(z_1)}{z_1^2}
\int_{0}^1 \text{d}\xi \int \frac{\text{d}
^{2}v_{\perp}}{(2\pi)^{2}}\frac{\text{d}^{2}v_{\perp}^{\prime
}}{(2\pi )^{2}} \frac{\text{d}^{2}u_{\perp}}{(2\pi)^{2}}
e^{-ik_{\perp }\cdot(v_{\perp}-v^{\prime }_{\perp})} \sum
\Psi^{\ast}_{j \to g g}(\xi, u_\perp)\Psi_{g\to gg}(\xi,
u_\perp)\{\text{Tr}U^\dagger(v^\prime_\perp)\nonumber \\
&&\times[\epsilon_{\perp}^{(1) } \cdot \partial
U(v^\prime_\perp)]U^\dagger(x_\perp)U(b_\perp) \text{Tr}
U^\dagger(b_\perp)U(x_\perp)-
\text{Tr}U^\dagger(v^\prime_\perp)[\epsilon_{\perp}^{(1) } \cdot
\partial U(v^\prime_\perp)]U^\dagger(b_\perp)U(x_\perp)
\text{Tr} U^\dagger(x_\perp)U(b_\perp)\}\nonumber \\
&&+ik^+\alpha_s\frac{D(z_1)}{z_1^2} \int_{0}^1 \text{d}\xi\int \frac{\text{d}%
^{2}v_{\perp}}{(2\pi)^{2}}\frac{\text{d}^{2}v_{\perp}^{\prime }}{(2\pi )^{2}}%
\frac{\text{d}^{2}u^\prime_{\perp}}{(2\pi)^{2}} e^{-ik_{\perp
}\cdot(v_{\perp}-v^{\prime }_{\perp})} \sum \Psi^{\ast}_{g\to g
g}(\xi, u^\prime_\perp)\Psi_{j \to gg}(\xi, u^\prime_\perp)
\{\text{Tr}U^\dagger(v_\perp)\nonumber \\
&&\times[\epsilon_{\perp}^{(1)\ast} \cdot \partial
U(v_\perp)]U^\dagger(x^\prime_\perp)U(b^\prime_\perp) \text{Tr}
U^\dagger(b^\prime_\perp)U(x^\prime_\perp) -
\text{Tr}U^\dagger(v_\perp)[\epsilon_{\perp}^{(1)\ast} \cdot
\partial
U(v_\perp)]U^\dagger(b^\prime_\perp)U(x^\prime_\perp) \text{Tr}
U^\dagger(x^\prime_\perp)U(b^\prime_\perp)\}.\nonumber
\\
\label{vr}
\end{eqnarray}
\begin{figure}[ht]
\begin{center}
\includegraphics[width=15cm]{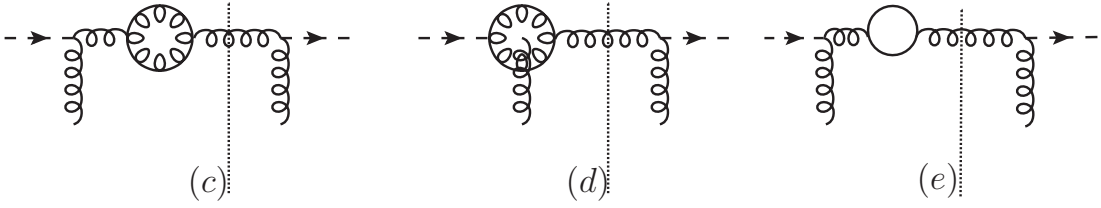}
\caption{Virtual loop diagrams in the  deep inelastic scattering
of current off a large nucleus at one-loop order.}\label{fig3}
\end{center}
\end{figure}
\end{widetext}
\newpage
 After lengthy calculations,
 where we only keep the leading power contribution,
we get the contribution of graph $(d)$ in Fig.~\ref{fig3} after
factoring out the leading order contribution,
\begin{eqnarray}
\frac{\alpha_sN_c}{\pi}
\left(-\frac{1}{\epsilon^2}+\frac{1}{\epsilon}\ln
\frac{Q^2}{\mu^2}-\frac{1}{2}\ln^2\frac{Q^2}{\mu^2}
+\frac{\pi^2}{12}\right). \label{vc}
\end{eqnarray}
We can see that $\pi^2/2$ in brackets is absent in Eq.~(\ref{vc})
 comparing to Eq.~($49$) of Ref.~\cite{Mueller:2013uf} since this process is space-like. Adding
Eq.~(\ref{rc}) and Eq.~(\ref{vc}) together, we can get the Sudakov
double logarithm term
\begin{equation}
- \frac{\alpha_s N_c}{2\pi}\ln^2 \frac{Q^2R_\perp^2}{c_0^2}.
\end{equation}
 \indent Next, we calculate the contributions of graph $(c)$ and graph $(e)$
  in Fig.~\ref{fig3}, the differential cross section of graphs $(c)$ in
  Fig.~\ref{fig3} can be cast into
\begin{eqnarray}
\frac{\text{d}\sigma_{c}^{jA\to h X}}{\sigma_0\text{d} y\text{d}^2
p_{h\perp}}&=&-\frac{1}{2}
 k^+ \alpha_s N_c\frac{D(z_1)}{z^2_1}\int_{0}^1 \text{d}\xi \notag \\ &&\times\int \frac{\text{d}%
^{2}v_{\perp}}{(2\pi)^{2}}\frac{\text{d}^{2}v_{\perp}^{\prime }}{(2\pi )^{2}}%
\frac{\text{d}^{2}u_{\perp}}{(2\pi)^{2}}  e^{-ik_{\perp
}\cdot(v_{\perp}-v^{\prime }_{\perp})} \notag \\ &&\times\sum
\Psi^{\ast}_{g\to g g}\left(\xi,
u_\perp\right)\Psi_{g\to gg}\left(\xi, u_\perp\right)\notag \\ &&\times S^{WW}(v_\perp,v^\prime_\perp),  \label{vl} % \notag \\
\end{eqnarray}
the differential cross section of graph $(e)$ in Fig.~\ref{fig3}
can be cast into
\begin{eqnarray}
\frac{\text{d}\sigma_{e}^{jA\to h X}}{\sigma_0\text{d} y\text{d}^2
p_{h\perp}}&=&- k^+ \alpha_s T_f N_f
\frac{D(z_1)}{z_1^2}\int_{0}^1 \text{d}\xi\notag \\ &&\times \int
\frac{\text{d}
^{2}v_{\perp}}{(2\pi)^{2}}\frac{\text{d}^{2}v_{\perp}^{\prime }}{(2\pi )^{2}}%
\frac{\text{d}^{2}u_{\perp}}{(2\pi)^{2}} e^{-ik_{\perp
}\cdot(v_{\perp}-v^{\prime }_{\perp})}\notag \\ &&\times \sum
\Psi^{\ast}_{g\to q \bar q}\left(\xi, u_\perp\right)\Psi_{g\to
q\bar q}\left(\xi, u_\perp\right)\notag \\ &&\times
 S^{WW}(v_\perp,v^\prime_\perp),  \label{vl2} % \notag \\
\end{eqnarray}
where $N_f$ is the number of quark flavors and $T_f =\frac{1}{2}$.
 The splitting function of $\Psi_{g\to q
\bar q}\left(\xi,u_\perp\right)$ can be found in
Ref.~\cite{Dominguez:2011wm, Chirilli:2011km} and the sum of the
splitting function is
\begin{eqnarray}
\sum&&\Psi^{\ast}_{g\to q \bar q}\left(\xi,
u_\perp\right)\Psi_{g\to q\bar q}\left(\xi, u_\perp\right)\notag
\\&&=\frac{2}{k^+} (2\pi)^2 \left[\xi^2+(1-\xi)^2\right]
\frac{1}{u_\perp^2}.
\end{eqnarray}
 The sum of graph $(c)$ and $(d)$ in Fig.~\ref{fig3} gives
\begin{eqnarray}
-\frac{\alpha_a
N_c}{\pi}\left[\beta_0\left(-\frac{1}{\epsilon_{\textrm{IR}}}+\ln
\frac{Q^2}{\mu^2}\right)+\beta_0\left(\frac{1}{\epsilon_{\textrm{UV}}}-\ln
\frac{Q^2}{\mu^2}\right)\right], \notag \\ \label{twov}
\end{eqnarray}
where $\beta_0 $ is $(11-2N_f)/12N_c$. We can see that there are
two divergences in Eq.~(\ref{twov}), infrared divergence and
Ultraviolet divergence, they should be subtracted in Sudakov
resummation. The term of Ultraviolet divergence
\begin{eqnarray}
   -\frac{\alpha_a
N_c}{\pi}\beta_0\left(\frac{1}{\epsilon_{\textrm{UV}}}-\ln
\frac{Q^2}{\mu^2}\right)\label{uv}
\end{eqnarray}
 is absorbed into the renormalization of the coupling
constant $\alpha_s$. The infrared divergence and the contribution
of Eq.~(\ref{partofRF}) are absorbed into the
 fragmentation function of the final state
hadron\cite{Chirilli:2011km} as follows
\begin{eqnarray}
  D_{h/g}(z_1,\mu)=D^{(0)}_{h/g}(z_1)-\frac{1}{\epsilon}\frac{\alpha_sN_c}{\pi}
  \int_{z_1}^1\frac{d\xi}{\xi}\mathcal{P}(\xi)_{gg}D_{h/g}(\frac{z_1}{\xi}),\notag \\
  \end{eqnarray}
where
\begin{equation}
\mathcal{P}_{gg}
(\xi)=\frac{\xi}{(1-\xi)_+}+\frac{1-\xi}{\xi}+\xi(1-\xi)+\beta_0
\delta(1-\xi).
\end{equation}
 After subtracting these two divergences, we can get the
single logarithm term of the Sudakov factor as follows
\begin{equation}
\frac{\alpha_sN_c}{\pi}\beta_0\ln\frac{Q^2R_\perp^2}{c_0^2},
\end{equation}
where we set the factorization scale $\mu^2=c_0^2/R_\perp^2$.
Adding the double and single logarithms terms together, we get the
Sudakov factor of DIS of a current off a large nucleus at one-loop
order
\begin{eqnarray}
{\cal S}_{\textrm{sud}}(Q^2,R_\perp^2)=\frac{\alpha_sN_c}{\pi}
\left(\beta_0 \ln \frac{Q^2R_\perp^2}{c_0^2}-\frac{1}{2}\ln^2
\frac{Q^2R_\perp^2}{c_0^2} \right).\notag \\ \label{oneloopsud}
\end{eqnarray}
\indent At the end of the day, assuming the exponentiation of
one-loop result, we can write down the differential cross section
of DIS of a gluonic current off a large nucleus at one-loop order
including Sudakov factor as
\begin{eqnarray}
\frac{d\sigma^{\rm ({\rm
resum})}}{\sigma_0dyd^2p_{h\perp}}|_{k^2_\perp\ll Q^2}&=&\frac{
D_{h/g}(z_1) }{z^2_1} \int \frac{d^2v_\perp
d^2v_\perp'}{(2\pi)^2}e^{ik_\perp\cdot(v_\perp- v^\prime_\perp)}
\notag \\&&\times e^{-{\cal S}_{\textrm{sud}}(Q^2,R_\perp^2)}
S^{WW}_{Y=\ln 1/x_g}(v_\perp,v_\perp').\notag \\
 \label{resum}
\end{eqnarray}
\section{ photon pair production in $pA$ collisions}
\begin{figure}[ht]
\begin{center}
\includegraphics[width=1.5in]{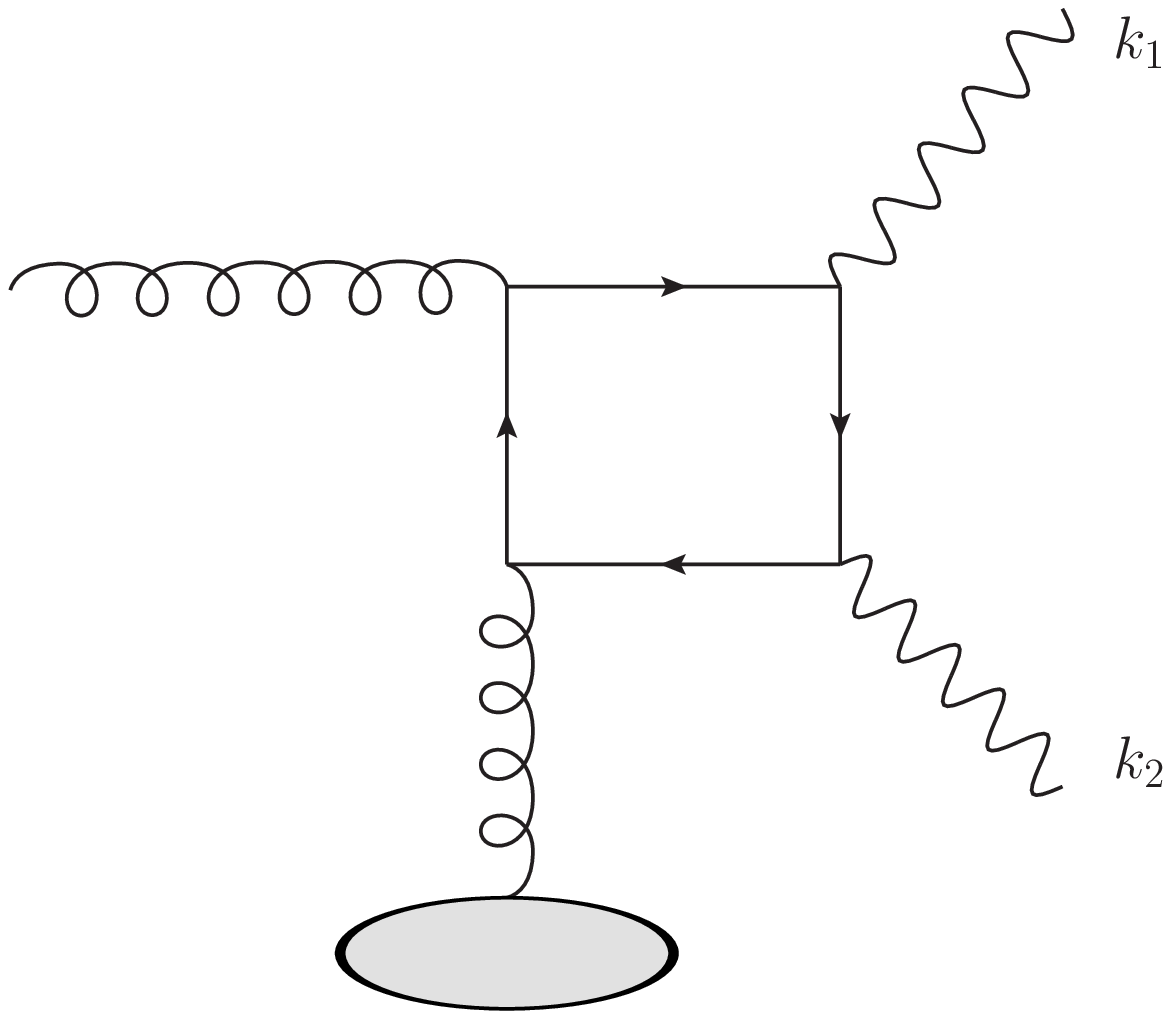}
\includegraphics[width=1.5in]{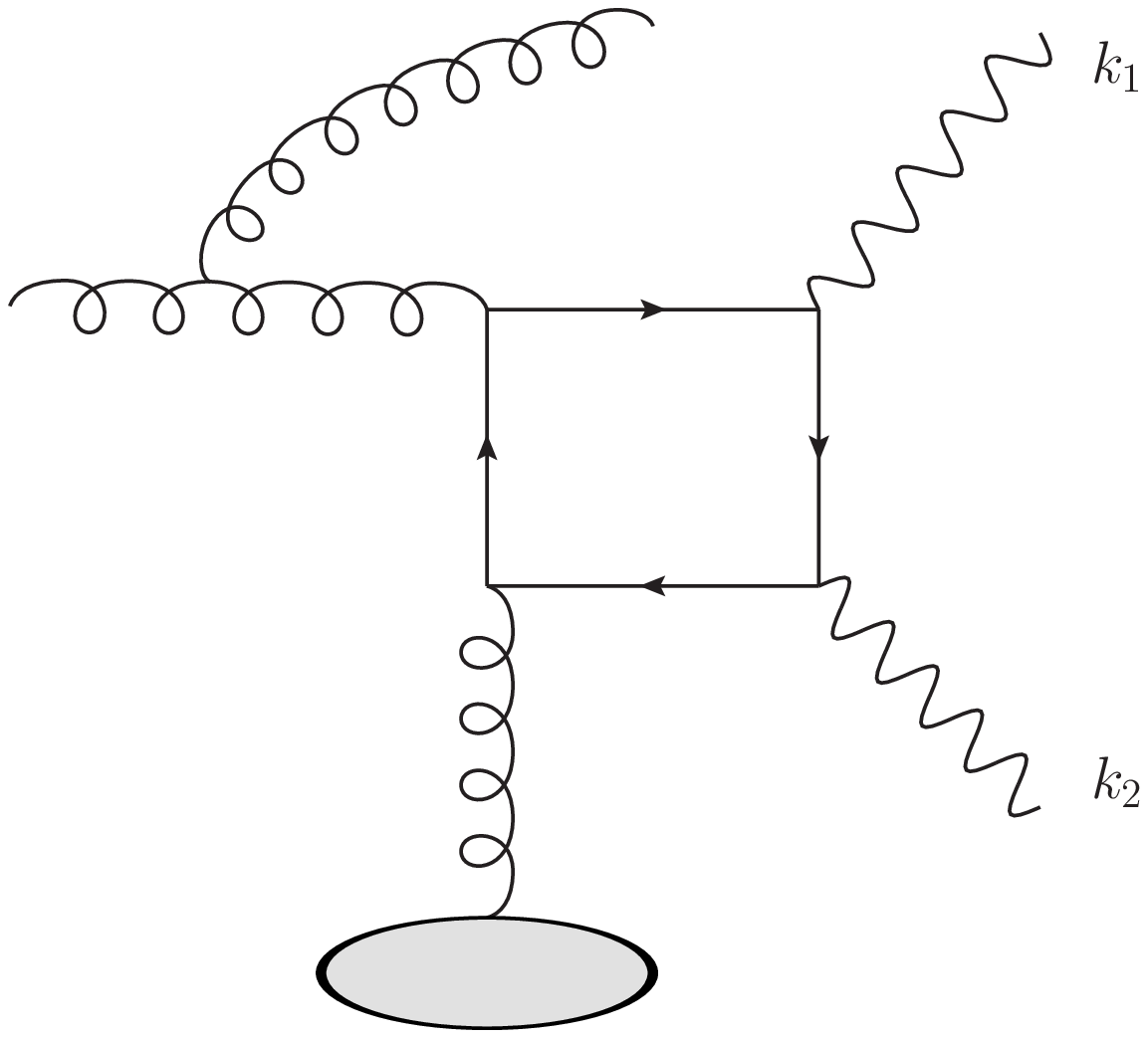}
\caption{Feynman diagrams of $pA\rightarrow \gamma\gamma+X$, the
left one is the leading-order contribution, the right one is the
real contribution at one-loop order.}\label{fig4}
\end{center}
\end{figure}

Two kinds of gluon distributions are introduced in
Ref.~\cite{Dominguez:2011wm}. The first gluon distribution is WW
gluon distribution which we have mentioned, the second one is
dipole gluon distribution, which is
 fourier transform of the dipole cross section. They are
different in many ways, the WW gluon distribution only contains
initial or final interaction, the dipole gluon distribution
contains both initial and final interaction. The WW gluon
distribution can be interpreted as the number density of gluons in
the light-cone gauge, but the diploe gluon distribution has no
such interpretation. They appear in different physics processes.\\
 \indent Let's consider the
process of $pA\rightarrow \gamma\gamma+X$, which is described in
Fig.~\ref{fig4}, The left graph in Fig.~\ref{fig4} is the feynman
diagram at leading-order , the right graph in Fig.~\ref{fig4} is
the feynman diagram at one-loop order, there are five other
feynman diagrams similar to each diagram, and they are omitted.
$k_1$ and $k_2$ are the momentum of the two observed photons,
$q_p$ is the momentum of the incoming gluon from proton, $p_A$ is
the momentum of the outgoing gluons from nucleus target. As we see
from the feynman diagram, there is only initial interaction in the
process of $pA\rightarrow\gamma\gamma+X$, thus, the involving
gluon distribution is WW gluon distribution. In this process, we
assume that the observed two photons are radiated back-to-back,
and have large transverse momentum.
 We define $q_\perp=|k_{1\perp}+k_{2\perp}|$ and
$P_\perp=|k_{1\perp}-k_{2\perp}|/2$, and assume that
$P_\perp\simeq|k_{1\perp}|\simeq|k_{2\perp}|\gg
q_\perp=|k_{1\perp}+k_{2\perp}|$. Thus, we should only keep the
contribution which is not suppressed by term of
$q_\perp^2/P_\perp^2$ when we are calculating the Sudakov factor for this process.\\
\indent Following the same strategy as in
Ref.~\cite{Dominguez:2011wm}, we can compute the Sudakov
contribution at the one-loop order. Eventually, the differential
cross section of $pA\to\gamma\gamma+X$ can be cast into
\begin{eqnarray}
  \frac{\text{d}\sigma^{pA\rightarrow \gamma\gamma+X}}
  {\text{d}y_1\text{d}y_2\text{d}^2P_{\perp}\text{d}^2q_{\perp}}&=&x_p f(x_p)\frac{2}{\pi\alpha_s}
 \frac{\text{d}\sigma^{gg\rightarrow
 \gamma\gamma}}{\text{d}t} \int\frac{d^2x_\perp}{(2\pi)^2}\frac{d^2x^\prime_\perp}{(2\pi)^2}\notag\\
 &&\times e^{iq_{\perp}\cdot(x_\perp-x^\prime_\perp)}S^{WW}(x_\perp,x_\perp')\notag \\
 &&\times e^{- {\cal S}_{\textrm{sud}}(P_\perp^2,R_\perp)}
\end{eqnarray}
 where $ y_1$ and $y_2$ are the rapidities of the two outgoing
photons,
 the $x_pf(x_p)$ is the
collinear gluon distribution of the proton.
 $\text{d}\sigma^{gg\rightarrow \gamma\gamma}/\text{d}t$ is
 the hard part of $gg\rightarrow \gamma\gamma
$\cite{Berger:1983yi, Dicus:1987fk, Bern:2001df}, the explicit
expression of the hard part can be found in Table 2 of
Ref.~\cite{Berger:1983yi}. The Sudakov double logarithm in this
process is the same as the one of Higgs production in $pA$
collisions, which is
\begin{equation}
  {\cal S}_{\textrm{sud}}(P_\perp^2,R_\perp)=\frac{\alpha_s N_c}{2\pi}
  \ln^2\frac{P_\perp^2R_\perp^2}{c^2_0},
\end{equation}
when we calculate the Sudakov double logarithms term.\\
 \indent In this process, we only keep
$x_g$ small. Since $x_px_gs\approx P_\perp^2$,  $x_p$ is large,
thus, $x_pf(x_p)$ can be obtained by database. If the differential
cross section of $pA\rightarrow\gamma\gamma+X$ is measured by
experiment, we can gain the information of the WW gluon
distribution $S^{WW}(x_\perp,x_\perp')$. It is an interesting way
to study the
WW gluon distribution at the LHC and RHIC.\\
\indent There is another channel of the photon pair production in
$pA$ collisions, the incoming particles from the proton and
nucleus are not gluons, but quark and antiquark, the hard part of
the channel is $\mathcal{H}_{q\bar{q}\rightarrow \gamma\gamma}$.
When the scattering energy is high, the quark and antiquark
distributions are much smaller than the gluon distributions, thus,
the channel of $q\bar{q}\rightarrow \gamma\gamma$ can be neglected
in photon pair production in $pA$ collisions.
\section{conclusion}
In summary, we consider two physics processes involving WW gluon
distribution. The first process is the DIS of current off a large
nucleus, and the second one is the photon pair production in $pA$
collisions. The WW gluon distribution in the
 DIS of current off a large nucleus only contains final
interaction, the WW gluon distribution of the
photon pair production in $pA$ collisions only contains initial interaction.\\
\indent Based on the leading power approximation, we have
calculated the Sudakov factor in DIS of
 gluoic current off a large nucleus at one-loop order.
 The contributions from many real and
virtual graphs are power suppressed under the leading power
approximation, and they are neglected in the calculation. In the
calculation, we find that there are various types of divergences
at one-loop order, the divergences must be separated out from the
Sudakov resummation. The collinear divergence is absorbed into
renormalization of the fragmentation function of the final state
hadron, the rapidity divergence is absorbed into the
renormalization of the WW gluon distribution, the UV divergence is
absorbed into the renormalization of coupling constant. After
subtracting the divergences,  we get Sudakov factor which include
double and single logarithms terms. Finally, we get differential
cross section of DIS of gluonic current off a large nucleus
including the
Sudakov factor at one-loop order.\\
 \indent We also
consider the process of photon pair production in $pA$ collisions,
where
 the final interaction is absent in this process. Thus, the
 involving
gluon distribution is the WW gluon distribution. Based on the
TMD-factorization, we get differential cross section expression
including the Sudakov factor at one-loop order. It is suggested
that if the differential cross section is measured at the LHC and
RHIC, the WW gluon distribution $S^{WW}(x_\perp,x_\perp')$ may be
extracted from the experimental data.
\begin{acknowledgments}
One of the authors Y.P. Xie thanks Dr. Bo-Wen ~Xiao for useful
comments and discussions. This work is supported in part by the
National Natural Science Foundation of China (Grants No.
11175220), the one Hundred Person Project (Grant No. Y101020BR0)
and Central China Normal University.
\end{acknowledgments}

\end{document}